\documentclass[twocolumn,preprintnumbers,secnumarabic,amsmath,amssymb,superscriptaddress,nofootinbib,floatfix]{revtex4}

\usepackage{graphicx}
\usepackage{dcolumn}
\usepackage{bm}

\usepackage{epsfig}
\usepackage{psfrag}
\usepackage{amsmath}

\def\beq{\begin{equation}}
\def\eeq{\end{equation}}
\def\be{\begin{equation}}
\def\ee{\end{equation}}
\def\bea{\begin{eqnarray}}
\def\eea{\end{eqnarray}}

\newcommand{\dif}{\,\mbox{d}}
\newcommand{\spec}{n_{\text{s}}}
\newcommand{\oW}{W^*}

\newcommand{\euler}{\text{e}}
\newcommand{\imag}{\text{i}}

\newcommand{\MP}{M_{\text{P}}}

\newcommand{\Ztwo}{\text{Z}_2}
\newcommand{\Zfour}{\text{Z}_4}
\newcommand{\Zsix}{\text{Z}_6}

\DeclareMathOperator{\diag}{diag}

\def\<{\left\langle}
\def\>{\right\rangle}

\begin{document}

\preprint{MPP-2009-18}

\medskip\

\title{SUGRA Hybrid Inflation with Shift Symmetry}
\author{Stefan Antusch\footnote{\tt
antusch@mppmu.mpg.de},
Koushik Dutta\footnote{\tt koushik@mppmu.mpg.de} and Philipp M. Kostka\footnote{\tt kostka@mppmu.mpg.de}}
\affiliation{Max-Planck-Institut f\"ur Physik (Werner-Heisenberg-Institut)\\
F\"ohringer Ring 6, D-80805 M\"unchen, Germany}

\begin{abstract}
We propose a solution to the $\eta$\,-\,problem in supergravity (SUGRA) hybrid inflation
using a Nambu-Goldstone-like shift symmetry within a new class of models. 
The flatness of the tree-level inflaton potential is ensured by  
shift symmetry invariance of the K\"ahler potential, while a small symmetry breaking term 
in the superpotential gives rise to a slope of the potential at loop-level.
In the proposed class of inflation models, potentially dangerous couplings between the 
inflaton and the moduli sector are avoided. 
We also discuss under which conditions the predicted spectral index can be in 
agreement with the best-fit-value of the latest WMAP observation $n_{\text{s}}\sim 0.96$,
corresponding to a hilltop-type inflaton potential at loop-level. 
\end{abstract}

\maketitle

\section{Introduction}
Inflation is considered as the standard paradigm for providing an 
explanation for the observed spatial flatness, homogeneity and isotropy on large 
scales as well as for the observed perturbations on smaller scales~\cite{Liddle:2000cg}.
From the point of view of particle physics, realizing inflation requires an extension 
of the Standard Model (SM). In order to connect inflation to particle physics motivated extensions 
of the SM, for example to Grand Unified Theories (GUTs) 
(see e.g.~\cite{Jeannerot:1997is}) or to models towards solving the flavor puzzle 
\cite{Antusch:2008gw}, hybrid-type inflation models
\cite{Linde:1991km,Copeland:1994vg,Linde:1997sj}
are particularly suitable. Since in general high energies close to the GUT scale are involved 
in models of this type, they are typically embedded in a supersymmetric (SUSY) framework. 

However, when trying to embed F-term hybrid inflation (or inflation in general) in local SUSY, 
i.e.\ supergravity theories, a severe problem arises: A general SUGRA theory seems to almost inevitably generate a large mass for the inflaton of the order of the Hubble scale $\mathcal{H}$ which results in a slow-roll parameter $\eta\sim 1$, spoiling inflation. This problem is referred to as the $\eta$\,-\,problem of SUGRA inflation \cite{Copeland:1994vg,Dine:1995uk}.
Several approaches to solve it have been discussed in the literature so far: For instance, since the fields involved in hybrid inflation typically have values well below the Planck scale $\MP$, one can expand the K\"ahler potential in powers of field values over $\MP$ and finally adjust (tune) the expansion parameters to protect the inflaton direction from obtaining large mass corrections (see e.g.\ \cite{Senoguz:2004ky,Antusch:2004hd,BasteroGil:2006cm}).
A different approach is to use a so-called Heisenberg symmetry~\cite{Gaillard:1995az, Antusch:2008pn},
which involves the inflaton as well as a modulus field. Here, the flatness of the inflaton potential is guaranteed by the symmetry. 
It has been shown in \cite{Antusch:2008pn} how 
this modulus field can be stabilized successfully during inflation in a novel class of 
models.   
The third approach towards solving the $\eta$\,-\,problem is to apply a Nambu-Goldstone-like shift symmetry~\cite{Kawasaki:2000yn,Yamaguchi:2000vm,Brax:2005jv} which can in principle protect the inflaton part of a complex scalar field against obtaining SUGRA mass corrections. However, it has turned out that in conventional (``standard'') models of SUSY hybrid inflation \cite{Copeland:1994vg, Dvali:1994ms}, the inflaton direction always develops a large tachyonic mass. To cure this instability, a coupling between the inflaton and the moduli sector has been considered, which however again leads to severe problems for realizing inflation as discussed in \cite{Brax:2006ay,Davis:2008sa}.

In this paper we propose a new model of SUGRA hybrid inflation where a Nambu-Goldstone-like shift symmetry is applied to guarantee the flatness of the inflaton potential. In contrast to the ``standard'' models of hybrid inflation in SUGRA, our model has the novel property of having $W = 0$ during inflation. Due to this property, potentially dangerous couplings between the inflaton and the moduli sector are absent. In addition, the problem of having a large tachyonic mass at tree-level for the inflaton direction is also cured in our model. We point out that this model forms part of a larger class of models which have the same desirable property. Furthermore, we note that in this class of models the right-handed sneutrino is a natural candidate for being the inflaton \cite{Antusch:2004hd}. 

Within the proposed class of models we address a second challenge to SUGRA models of hybrid inflation, namely the consistency between the predicted spectral index and the latest WMAP results \cite{Komatsu:2008hk,Hinshaw:2008kr}, which favor a value of $n_{\text{s}}\sim 0.96$ for small tensor-to-scalar ratio $r$. In fact, ``standard'' SUSY hybrid inflation models with flat inflaton potential at tree-level predict a spectral index above $n_{\text{s}}\sim 0.98$~\cite{Dvali:1994ms}. Since this is slightly outside the $1\sigma$\,-\,range given by WMAP, it raises the question whether a lower spectral index can be realized. 
In this context, an expansion of the K\"ahler potential has been considered and it has been shown that a lowering of $n_{\text{s}}$ to $n_{\text{s}}\sim 0.96$ can be achieved \cite{BasteroGil:2006cm,Pallis:2009pq} by a small term which, however, already introduces a curvature of the potential at tree-level. Another possibility, pointed out in 
\cite{Battye:2006pk}, is that a cosmic string component could allow for a more blue-tilted power spectrum of the Cosmic Microwave Background (CMB). Other approaches suggest, e.g., \ to involve field values 
well above the Planck-scale~\cite{Clesse:2008pf}.

Here we propose and analyze a new possibility to lower the spectral index $n_{\text{s}}$ in SUGRA hybrid inflation models based on a new term in the K\"ahler potential involving the waterfall field and the field which gives rise to the vacuum energy by its F-term. While the inflaton potential remains flat at tree-level, the new term changes the quantum loop corrections and therefore the form of the scalar potential. We show that without any tuning of parameters, the spectral index can be lowered and the WMAP best-fit value for $n_{\text{s}}$ can be realized.

The paper is organized as follows: 
In section~\ref{The Model}, we
describe the model given in terms of the
superpotential and K\"ahler potential and
the chiral superfield content. Additionally,
the shift symmetry in the inflaton sector 
is introduced.
Section~\ref{Effective Potential} is dedicated
to the investigation of the loop-corrected
scalar potential and the tree-level mass spectrum 
to show that all other directions 
in field space are stable during 
inflation. Section~\ref{Predictions} contains 
the calculation of observables obtained 
from the effective potential and finally, in 
section~\ref{Summary and Conclusions}, 
we summarize and present our conclusions.

\section{The Model}\label{The Model}
To solve the $\eta$-problem in SUGRA hybrid inflation using a Nambu-Goldstone-like shift symmetry, we propose the following minimal setup described by the superpotential 
\begin{equation}\label{tribridsuperpotential}
W=\kappa\,S\left(H^2-M^2\right)+\frac{\lambda}{M_{*}}N^2H^2\,,
\end{equation}
with $N$ being the superfield which contains the inflaton, and the K\"ahler potential
\begin{equation}\label{kaehlerpotential}
\begin{split}
 K=\,&|H|^2+|S|^2+\frac{1}{2}\left(N+N^*\right)^2+
\frac{\kappa_{H}}{\Lambda^2}\,|H|^4\\
&+\frac{\kappa_{S}}{\Lambda^2}\,|S|^4
+\frac{\kappa_{N}}{4\,\Lambda^2}\,\left(N+N^*\right)^4
 +\frac{\kappa_{SH}}{\Lambda^2}\,|S|^2|H|^2\\
&+\frac{\kappa_{SN}}{2\,\Lambda^2}\,|S|^2\left(N+N^*\right)^2\\
&+\frac{\kappa_{HN}}{2\,\Lambda^2}\,|H|^2\left(N+N^*\right)^2+\ldots\,.
 \end{split}
 \end{equation}
The gauge singlet supermultiplet $N$ contains  
the slow-rolling inflaton condensate as the pseudoscalar 
component which will turn out to be a tree-level flat 
direction. $S$ is a static, so-called driving 
field which is fixed at zero during inflation and only contributes the large vacuum energy density during 
inflation by its F-term. $H$ is the waterfall field which also
stays at zero during inflation, but gets destabilized when the inflaton
reaches a critical value. Both $S$ and $H$ are also gauge singlets in our minimal model.  
The form of the superpotential is motivated by the model of \cite{Antusch:2004hd} and the classes of models considered in \cite{Antusch:2008pn}, where the right-handed sneutrino has been identified as the inflaton field.
It has the desirable property of having vanishing superpotential $W=0$
both during and after inflation, and all the derivatives vanish,
except for the one $W_{S}\ne 0$, as we will discuss below.

As an aside, we would like to note at this point that in ``conventional'' SUGRA models of hybrid inflation the superpotential consists of only the first term $W=\kappa\,S\left(H^2-M^2\right)$ in Eq.~(\ref{tribridsuperpotential}) and the superfield $S$ contains the inflaton. The K\"ahler potential for $S$ is chosen to be minimal, i.e.\ $K = |S|^2$. Large SUGRA contributions to the inflaton mass get cancelled by this special choice of $K$, however not protected by a fundamental symmetry. 
To use a shift symmetric K\"ahler potential in combination with the above superpotential to protect the flatness of the inflaton direction, the scalar potential would always contain a tachyonic direction for one of the components of the complex inflaton field. 
One possibility to stabilize it could be the introduction of
a coupling to another sector (i.e. to a modulus field). 
However, as has been argued in \cite{Brax:2006ay}, such a coupling is very problematic for successfully realizing inflation. On the contrary, the model we propose is free of these problems.

Turning to the K\"ahler potential, it is symmetric under the ``shift'' of $N$ by 
\begin{equation}\label{shiftsymmetry}
N\,\rightarrow \,N+\imag\,\mu\,,
\end{equation}
where $\mu$ is an arbitrary dimensionful parameter.
Shift symmetry invariance of the K\"ahler potential requires it to be a function 
$K(N,N^*)=K(N+N^*)$.\footnote{Shift symmetry in the context of natural inflation with Pseudo-Nambu-Goldstone-Bosons has been mentioned in~\cite{Freese:1990rb}. 
It also arises in string theory and plays an important role in string inflation model building, as described e.g.\ in Ref.~\cite{Hsu:2004hi}.  } 
For generality, and since all field values are well below the Planck scale,   
we use an expansion in the absolute values squared 
of the fields $H$ and $S$ and in even powers of
$(N+N^*)/\sqrt{2}$, where higher order operators are suppressed 
by suitable powers of some scale $\Lambda$. The scale $\Lambda$ can also be viewed as the generation scale of the effective operators, and can well be below the Planck scale $M_P$.

The fact that the shift symmetry of the K\"ahler potential provides a solution to the $\eta$-problem in SUGRA inflation can be seen by looking at the full (F-term) scalar potential given by
\footnote{Here we use a convention in which we set the reduced 
Planck mass $\MP\simeq 2.4\times 10^{18}\,\text{GeV}$ to unity.}
\begin{equation} \label{Fterm}
V_{\text{F}}=\text{e}^K\left[K^{i\bar{j}}\,
\mathcal{D}_{i}W\,\mathcal{D}_{\bar{j}}\oW
 - 3|W|^2\right]\,,
\end{equation}
where the definition 
\begin{equation}
\mathcal{D}_{i}W:=W_{i}+K_{i}\,W
\end{equation}
has been used. The indices $i, j$ denote different scalar fields and the lower indices on the superpotential or K\"ahler potential denote the derivatives with respect to the scalar component of the chiral superfield or their conjugate where a bar is involved. The inverse K\"ahler metric is defined as $K^{i \bar j} = K^{-1}_{i\bar j}$.

The $\eta$-problem is the tendency that in SUGRA an inflaton mass of the order of the Hubble
scale is typically generated, which spoils inflation. 
Because of the shift symmetry the exponential SUGRA contribution in Eq.~\eqref{Fterm} is independent of the imaginary part of the $N$ field, identified as the inflaton direction. This solves the usual $\eta$-problem in SUGRA inflation. In conventional hybrid inflation models $W \neq 0$ during inflation and this can also induce a large tachyonic mass for the inflaton field at tree-level from the $-3|W|^2$ term in Eq.~\eqref{Fterm}. This problem is automatically avoided in our setup due to the special property of having $W=0$ during inflation. Furthermore, this property also eliminates certain couplings between the moduli sector and the inflaton which have turned out to be problematic for successfully realizing inflation \cite{Brax:2006ay}.

The K\"ahler metric is given by the second derivatives 
w.r.t.\ the fields and their conjugates. In the limit $S,H\rightarrow 0$,
it diagonalizes and in $(H,S,N)$-basis is given by
\begin{equation}\label{kaehlermetric}
K_{i\bar{j}}=\delta_{i\bar{j}}+\diag\left(\frac{\kappa_{HN}}{\Lambda^2}\,n^2_{\text{R}}\,,\,
\frac{\kappa_{SN}}{\Lambda^2}\,n^2_{\text{R}}\,,\,
6\frac{\kappa_{N}}{\Lambda^2}\,n^2_{\text{R}}\right)\,.
\end{equation}
With $n_{\text{R}}=0$, we recover canonically 
normalized kinetic terms for 
$N=(n_{\text{R}}+\imag\,n_{\text{I}})/\sqrt{2}$,
$H=(h_{\text{R}}+\imag\,h_{\text{I}})/\sqrt{2}$ and
$S=s/\sqrt{2}$.
The shift symmetry, together with the form of the superpotential, protects 
the imaginary part of $N$ from obtaining any mass term at tree-level.

Finally, the shift symmetry is slightly broken by the effective operator $\frac{\lambda}{M_{*}}N^2H^2$ of the superpotential in Eq.~(\ref{tribridsuperpotential}). While the inflaton potential is still flat at tree-level, this term gives rise to a slope of the inflaton potential at loop-level, as will be discussed in section~\ref{Effective Potential}. We note that the minimal superpotential of Eq.~(1) may be justified by introducing additional symmetries and spurion fields, as outlined in the Appendix. 

We would also like to note that the above example model can be generalized to a whole class of models, along the line of \cite{Antusch:2008pn}, by allowing for variants of the form of the superpotential.  
Furthermore, a specific (shift symmetric) K\"ahler potential may be chosen instead of the expansion we have used here for generality. The main point we would like to emphasize is that in our class of models with $W=W_N=0$ (but $W_S \not=0$) during inflation, a shift symmetry naturally solves the $\eta$-problem, and that in the same class of models several problems of earlier approaches, related to tachyons or to dangerous couplings to moduli fields, are automatically avoided.

\section{Effective Potential}\label{Effective Potential}
The full F-term scalar potential is given by Eq.~\eqref{Fterm}. 
In the following, we will assume that the fields $s,h_{\text{R}},h_{\text{I}}$ and $n_{\text{R}}$ have already settled to their minima at $s=h_{\text{R}}=h_{\text{I}}=n_{\text{R}}=0$. 
This is justified because, as we will show below, the fields can easily have masses larger than the Hubble scale $\mathcal{H}$. 
From Eq.~\eqref{kaehlermetric} we see that, 
in the minimum, we have $K_{i\bar{j}}=\delta_{i\bar{j}}$
such that all fields are already canonically normalized.
The vacuum energy density is then given by the
(F-term) tree-level potential which reads
\begin{equation}
V_{\text{tree}}=\kappa^2\,M^4\simeq 3\,\mathcal{H}^2\,,
\end{equation}
and is flat in $n_{\text{I}}$\,-\,direction.

Using the SuperCosmology code~\cite{Kallosh:2004rs}, 
the scalar and pseudoscalar mass matrices are calculated
to be diagonal and the mass spectrum is given by
\begin{equation}\label{scalarmasses}
\begin{split}
m^2_{n_{\text{I}}}&=0\,,\\
m^2_{n_{\text{R}}}&=2\,\kappa^2M^4\left(1-\frac{\kappa_{SN}}{\Lambda^2}\right)\,,\\
m^2_{s}&=-\,\frac{4\,\kappa_{S}}{\Lambda^2}\,\kappa^2M^4\,.
\end{split}
\end{equation}
The directions in field space different from the inflaton
are stable provided that their masses are larger than
the Hubble scale. This requirement leads to the 
constraints $\kappa_{SN}/\Lambda^2<5/6$ and $\kappa_{S}/\Lambda^2<-1/12$.

For the fermions, we calculate the SUGRA mass matrix from
\begin{equation}\label{fermionmassmatrix}
\begin{split}
\left(\mathcal{M}_{\text{F}}\right)_{ij}=&
\,\euler^{K/2}\big(W_{ij}+K_{ij}\,W+K_{i}\,W_{j}\\
&+K_{j}\,W_{i}
+K_{i}\,K_{j}\,W
-{K^{k\bar{l}}}\,K_{ij\bar{l}}\,\mathcal{D}_{k}W\big)\,.
\end{split}
\end{equation}
This is also diagonal in the minimum and 
has only one non-vanishing eigenvalue during inflation.
The fermion mass matrix squared in $(H,S,N)$-basis
is obtained to be 
\begin{equation}\label{fermionmassmatrix}
\left(\mathcal{M}_{\text{F}}\right)^2_{ij}=
\diag\left(\frac{\lambda^2}{M_{*}^2}\,n^4_{\text{I}}\,,\,0\,,\,0\right)\,.
\end{equation}

Since for both the scalars and pseudoscalars, as well as for the
fermions, only the waterfall sector $H$-components contribute
$n_{\text{I}}$\,-\,dependent masses, we only consider their
contributions to the one-loop potential. 
We denote their masses by
\begin{equation}\label{loopmasses}
\begin{split}
m^2_{\text{S}}&=2\,\kappa^2M^2\,\left[x+M^2\left(1-\frac{\kappa_{SH}}{\Lambda^2}\right)/2-1\right]\,,\\
m^2_{\text{P}}&=2\,\kappa^2M^2\,\left[x+M^2\left(1-\frac{\kappa_{SH}}{\Lambda^2}\right)/2+1\right]\,,\\
m^2_{\text{F}}&=2\,\kappa^2M^2\,x\,,
\end{split}
\end{equation}
where we have defined the new variable
\begin{equation}
x\equiv\left(\frac{\lambda}{\kappa}\right)^2\frac{n_{\text{I}}^4}{2\,(MM_{*})^2}\,.
\end{equation}
Hence, the critical value of $n_{\text{I}}$ at which
the waterfall field destabilizes can be calculated from
$m^2_{\text{S}}=0$ and is given by
\begin{equation}
(n^{\text{c}}_{\text{I}})^2=
\frac{\kappa}{\lambda}\,
(MM_{*})\,\sqrt{2+M^2\left(\frac{\kappa_{SH}}{\Lambda^2}-1\right)}\,.
\end{equation}

The Coleman-Weinberg one-loop contribution \cite{Coleman:1973jx}
to the effective 
potential using the squared masses in Eq.~\eqref{loopmasses}
reads
\begin{equation}\label{loop Potential2}
\begin{split}
V_{\text{loop}}=
\,&\frac{1}{64\,\pi^2}\,\Big[
m^4_{\text{S}}\left(\ln\frac{m^2_{\text{S}}}{Q^2}-\frac{3}{2}\right)\\
&+m^4_{\text{P}}\left(\ln\frac{m^2_{\text{P}}}{Q^2}-\frac{3}{2}\right)
-2\,m^4_{\text{F}}\left(\ln\frac{m^2_{\text{F}}}{Q^2}-\frac{3}{2}\right)\Big]\,,
\end{split}
\end{equation}
where $Q$ is the renormalization scale which we fix to $Q=m_{\text{F}}/\sqrt{x}=\sqrt{2}\,\kappa\,M$.

\begin{figure}[!h]
\psfrag{nc}{\hspace{0.06cm}$n^{\text{c}}_{\text{I}}$}
\psfrag{ne}{\hspace{0.06cm}$n^{\text{e}}_{\text{I}}$}
\psfrag{pot}{\hspace{-1cm}$V\,[10^{-13}\,\MP^4]$}
\psfrag{Vloop}{\hspace{-0.8cm}$V_{\text{tree}}+V_{\text{loop}}$}
\psfrag{Vtree}{$V_{\text{tree}}$}
\psfrag{ni}{\hspace{-0.1cm}$n_{\text{I}}\,[\MP]$} \center
\includegraphics[scale=0.8]{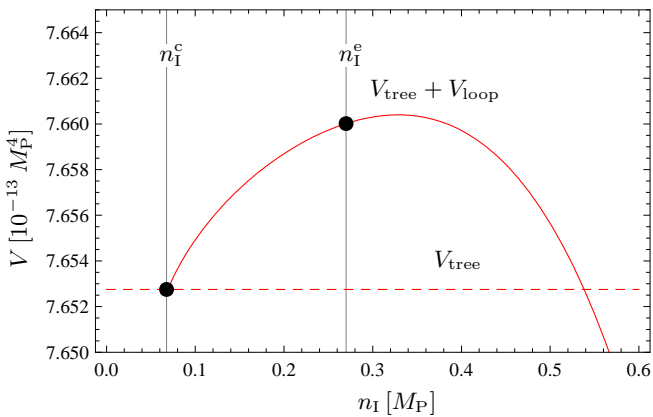}
\caption{\label{shiftpot}Graphical illustration of the one-loop
effective potential for $n_{\text{I}}$ with typical values of the
field 60 e-folds before the end of inflation
$n^{\text{e}}_{\text{I}}$ and at the critical value
$n^{\text{c}}_{\text{I}}$ where inflation ends.
The parameter choice in this example was
$\kappa=0.12$, $\lambda=0.1$, $(\kappa_{SH}\,\MP^2/\Lambda^2)=40$ and $M=0.0027\MP$.}
\end{figure}

The complete effective scalar potential 
we consider thus reads
\begin{equation}\label{effectivepotential}
V_{\text{eff}}(n_{\text{I}})=V_{\text{tree}}+V_{\text{loop}}(n_{\text{I}})\,.
\end{equation}
For a particular choice of parameters, we
have plotted the potential in Fig.~\ref{shiftpot}.
Note, that the shape of the potential is of 
hilltop-type, generated purely by radiative 
corrections.\footnote{It has recently been argued
in Ref.~\cite{Rehman:2009wv}
that such hilltop-type potentials can also arise
in non-supersymmetric hybrid inflation
when the inflaton is coupled to the right-handed neutrinos.}
In order to achieve at least 60 e-folds of inflation, 
the initial condition for the inflaton field is that it has to start between
$n^{\text{e}}_{\text{I}}$, about $60$ e-folds before the end of inflation, 
and the top of the hill (i.e.\ the local maximum of the potential). 
We note that, although certain initial conditions have to be satisfied for inflation to start, 
this does not impose any tuning since the relative amount
of field space where the initial conditions are suitable is quite extended in our model, about 
$20\%$ (c.f.\ Fig.~\ref{shiftpot}).

\section{Predictions}\label{Predictions}
\begin{figure}[!h]
\psfrag{tens}{\footnotesize\hspace{-0.4cm}$r\,[10^{-5}]$} 
\psfrag{ns}{\footnotesize$n_{\text{s}}$}
\psfrag{scale}{\footnotesize\hspace{-0.6cm}$M/\MP\,[10^{-3}]$}
\psfrag{kappaSH}{\hspace{-0.5cm}\footnotesize$\kappa_{SH}\,\MP^2/\Lambda^2$}
\psfrag{exit}{\hspace{-0.3cm}\footnotesize$n^{\text{e}}_{\text{I}}/\MP$} 
\psfrag{dns}{\hspace{-0.7cm}\footnotesize$\text{d}n_{\text{s}}/\text{d}\ln{k}\,[10^{-4}]$}
\center
\includegraphics[scale=0.7]{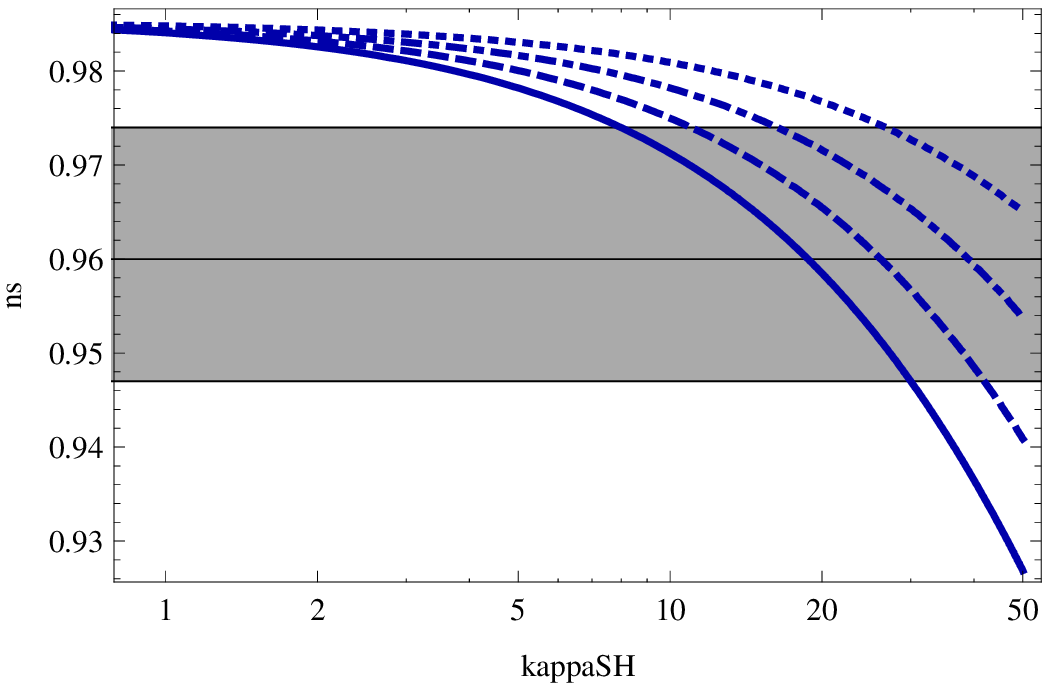}\\
\includegraphics[scale=0.7]{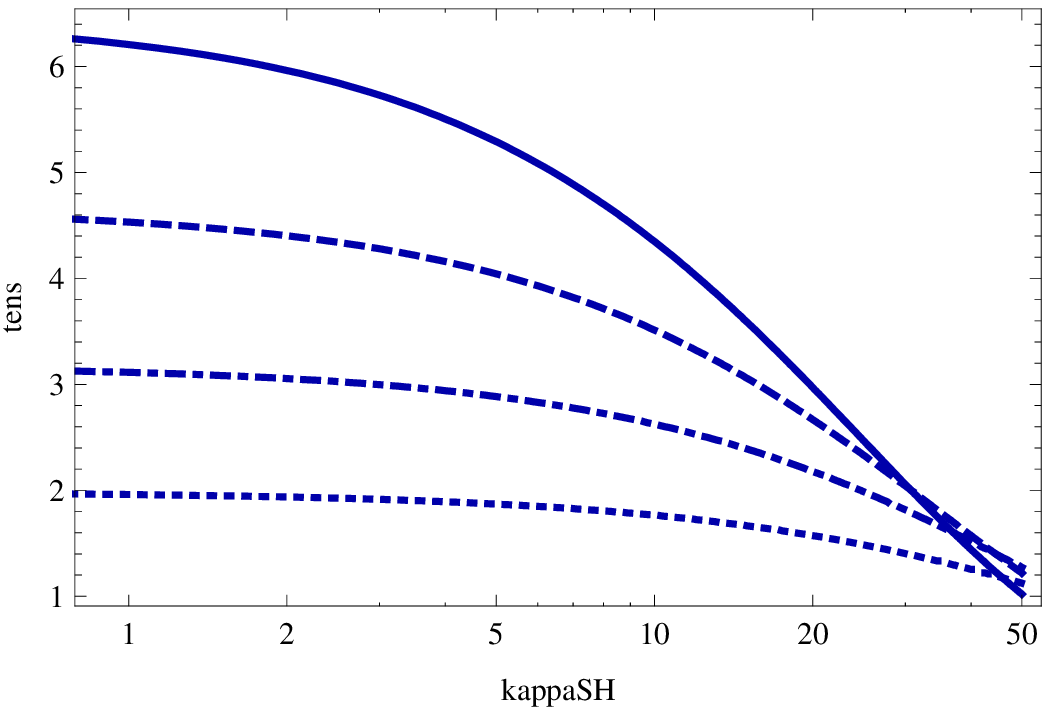}
\caption{\label{paramspace}Observables $n_{\text{s}}$ and $r$
for 
$\lambda=0.1$ and
$\kappa=0.14$ (full line), $\kappa=0.12$ (dashed line),
$\kappa=0.1$ (dotted dashed line), $\kappa=0.08$ (dotted line)
depending on $(\kappa_{SH}\,\MP^2/\Lambda^2)$.
The shaded region in the first plot highlights the latest WMAP
$1\sigma$\,-\,range of the spectral index.}
\end{figure}

\begin{figure}[!h]
\psfrag{tens}{\footnotesize\hspace{-0.4cm}$r\,[10^{-5}]$} 
\psfrag{ns}{\footnotesize$n_{\text{s}}$}
\psfrag{scale}{\footnotesize\hspace{-0.6cm}$M/\MP\,[10^{-3}]$}
\psfrag{kappaSH}{\hspace{-0.5cm}\footnotesize$\kappa_{SH}\,\MP^2/\Lambda^2$}
\psfrag{exit}{\hspace{-0.3cm}\footnotesize$n^{\text{e}}_{\text{I}}/\MP$} 
\psfrag{dns}{\hspace{-0.7cm}\footnotesize$\text{d}n_{\text{s}}/\text{d}\ln{k}\,[10^{-4}]$}
\center
\includegraphics[scale=0.7]{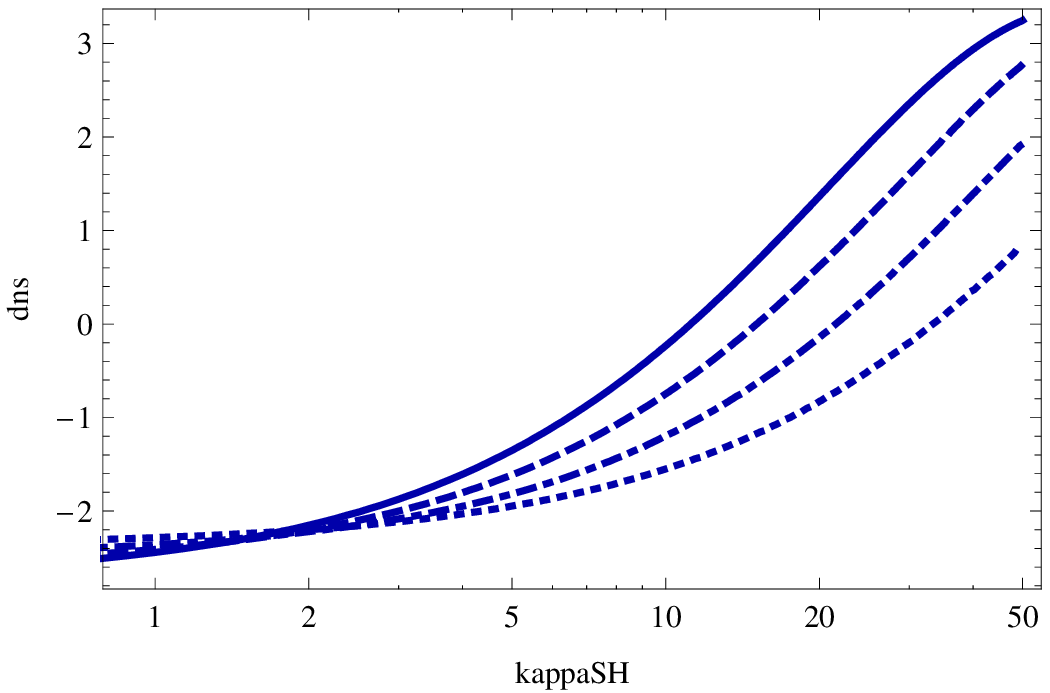}\\
\includegraphics[scale=0.7]{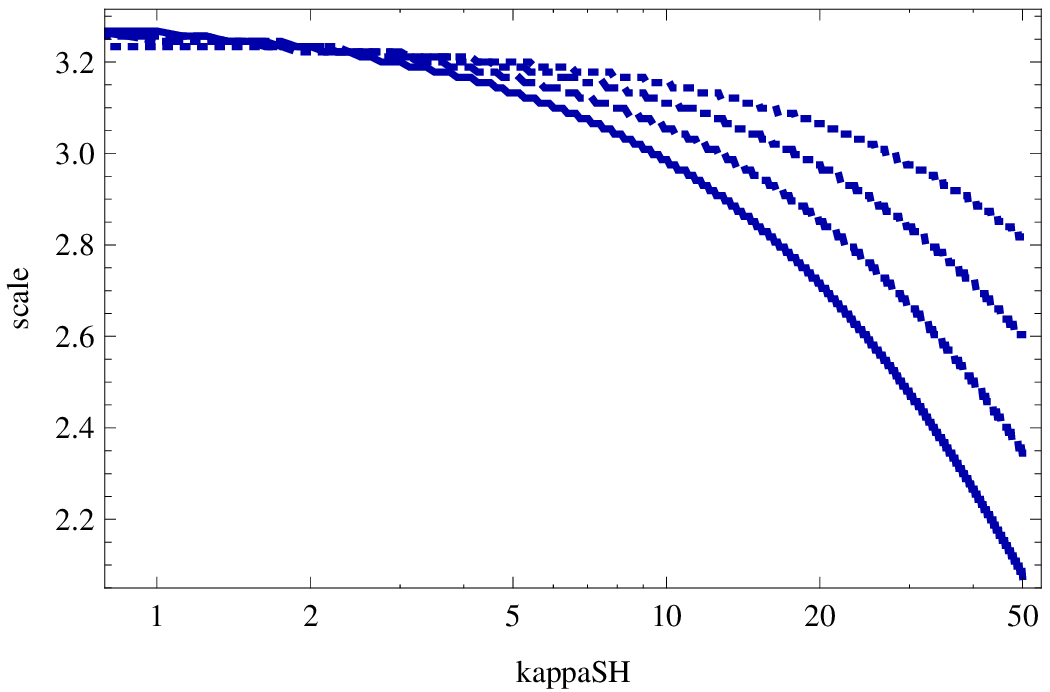}\\
\caption{\label{paramspace2} Running of the 
spectral index $\text{d}n_{\text{s}}/\text{d}\ln{k}$
and the scale $M$ for the same choice of parameters as in Fig.~\ref{paramspace}.}
\end{figure}

In order to make predictions for the observable quantities 
in the CMB radiation, we have to integrate the 
equations of motion (EOMs) for $n_{\text{I}}$
from $n^{\text{c}}_{\text{I}}$ to the value $n^{\text{e}}_{\text{I}}$ where the relevant 
scales leave the horizon, 
i.e.\ about $60$ e-folds before the end of inflation.
In slow-roll approximation, the EOMs read
\begin{equation}
3\,H\,\dot{n}_{\text{I}}(t)=-V'(n_{\text{I}})\,.
\end{equation}
The slow-roll parameters are defined as
\begin{equation}
\epsilon=\frac{1}{2}\left(\frac{V'}{V}\right)^2\,,\quad
\eta=\left(\frac{V''}{V}\right)\,,\quad
\xi^2=\left(\frac{V'V'''}{V^2}\right)\,.
\end{equation}
In terms of the slow-roll parameters, the scalar spectral index $n_{\text{s}}$, 
the tensor-to-scalar ratio $r$ and the running of the scalar spectral index ${\text{d}n_{\text{s}}}/{\text{d}\ln{k}}$ are given by
\begin{equation}\label{observables}
\begin{split}
n_{\text{s}}&\simeq 1-6\,\epsilon+2\,\eta\,,\\
r&\simeq16\,\epsilon\,,\\
\frac{\text{d}n_{\text{s}}}{\text{d}\ln{k}}&\simeq
16\,\epsilon\,\eta-24\,\epsilon^2-2\,\xi^2\,.
\end{split}
\end{equation}
In addition, the amplitude of the curvature perturbations can be
obtained from
\begin{equation}\label{power}
P^{1/2}_{\mathcal{R}}=\frac{1}{2\sqrt{3}\,\pi}\,
\frac{V^{3/2}}{|V'|}\,,
\end{equation}
where in our case $V=V_{\text{eff}}$ from Eq.~\eqref{effectivepotential}.
All of the above quantities have to be evaluated at the field value $n^{\text{e}}_{\text{I}}$, about 
$60$ e-folds before the end of inflation.
The scale $M$ is normalized to fit the observation
of the amplitude~\eqref{power}, which at $68$\% CL
is measured to be $P^{1/2}_{\mathcal{R}}\simeq(5.0\pm0.1)\cdot 10^{-5}$.
Fixing the parameters $M_{*}=\MP=1$ and $Q$ as above,
we vary $\kappa_{SH}$ for different values of $\kappa$ while
keeping fixed $\lambda=0.1$. 
The results for the observables calculated in this way are 
displayed in Figs.~\ref{paramspace} and~\ref{paramspace2}.

Note, that for a relatively wide and generic range
of the parameter $\kappa_{SH}$, the predicted 
spectral index $\spec$ lies within the $1\sigma$-range
$n_{\text{s}}=0.960^{+0.014}_{-0.013}$
as taken from the latest WMAP 
data~\cite{Komatsu:2008hk,Hinshaw:2008kr}.
The tensor-to-scalar ratio is very small $r\sim\mathcal{O}(10^{-5})$
as typical for hybrid and small-field models of inflation and 
easily satisfies the observational upper bound $r<0.2$ (95\% CL).
For a running of the spectral index, the current observational 
evidence is still not considered as significant, $\dif n_{\text{s}}/\dif\ln{k}=-0.032^{+0.021}_{-0.020}$ (68\% CL).
Our prediction $\dif n_{\text{s}}/\dif\ln{k}\sim\mathcal{O}(\pm10^{-4})$ 
does not lie within this range, which we do not consider
as a problem for our model due to the low accuracy of the measurement.
We note that the scale $M$ obtained from the COBE-normalization 
for the amplitude of the power spectrum 
predicts $\langle H\rangle\sim\mathcal{O}(10^{-3}\MP)$,
close to the GUT scale. 
In this simplest type of effective single-field model,
we do not expect large non-Gaussianities.

\section{Summary and Conclusions}\label{Summary and Conclusions}
In this paper, we have proposed a solution to the $\eta$\,-\,problem in 
supergravity (SUGRA) hybrid inflation using a Nambu-Goldstone-like shift 
symmetry within a new class of models (see Eqs.~(\ref{tribridsuperpotential}),  
(\ref{kaehlerpotential}) and generalizations discussed in the text). 
The flatness of the tree-level inflaton potential is ensured by  
shift symmetry invariance of the K\"ahler potential, while a small symmetry 
breaking term in the superpotential gives rise to a slope of the potential at loop-level.
Due to the special property of the class of models that during inflation $W = W_N = 0$, with $N$ 
being the superfield which contains the inflaton, the appearance of tachyonic directions in the 
inflaton potential is avoided and certain dangerous couplings between the moduli 
sector and the inflaton are eliminated.
The right-handed sneutrino provides an interesting particle physics candidate for 
being the inflaton. 
We have furthermore proposed a new possibility to lower the spectral index $n_{\text{s}}$ 
in SUGRA hybrid inflation models based on a new term in the K\"ahler potential 
involving the waterfall field and the field which gives rise to the vacuum energy 
by its F-term. 
While the inflaton potential remains flat at tree-level, the new term changes 
the quantum loop corrections and therefore the form of the scalar potential can become of
hilltop-type. This leads to a reduced $n_{\text{s}}$ compared to the generic prediction of 
$n_{\text{s}} \gtrsim 0.98$ in ``conventional'' SUSY models of hybrid inflation. 
We have shown that with the new term in the K\"ahler potential the spectral index can be lowered 
without any tuning of parameters 
and the WMAP best-fit value $n_{\text{s}}\sim 0.96$ can be realized.

\section*{Acknowledgments}
The authors would like to thank Steve F. King and Mar Bastero-Gil for discussions and comments on the draft. 
The authors acknowledge partial support by the DFG cluster of excellence 
``Origin and Structure of the Universe''.

\section*{Appendix: Symmetries and the Form of the Superpotential}
In this appendix we would like to outline how the simple superpotential and K\"ahler potential of Eqs.~(1) and (2) could be realized in an explicit model and, in particular, how additional unwanted terms (for instance a superpotential term $\sim N^2$) can be avoided. 

We therefore assume three additional superfields $X$, $Y$ and $Z$ which are heavy and, at the time of inflation, have already acquired VEVs in their bosonic components. 
We next impose the discrete symmetries $\Ztwo$, $\Zfour$, $\Zsix$
as well as an R-symmetry under which the superpotential carries charge $1$.
The field content of the toy model (which should be understood as a ``proof of existence'' only) together with the assigned charges are displayed in table \ref{symmetries}.

\begin{table}[!h]
\begin{tabular}{|c|c|c|c|c|c|c|}
\hline
&$H$&$S$&$N$&$X$&$Y$&$Z$\\\hline
$\Ztwo$&$-$&$+$&$-$&$+$&$+$&$+$\\\hline
$\Zfour$&$0$&$2$&$0$&$1$&$1$&$0$\\\hline
$\Zsix$&$2$&$0$&$1$&$4$&$0$&$5$\\\hline
R&$1/2$&$0$&$0$&$0$&$1/2$&$0$\\\hline
\end{tabular}
\caption{\label{symmetries} Superfield content and imposed symmetries.}
\end{table}

With these symmetries, the following terms in the superpotential (up to operators of dimension 6) and K\"ahler potential are allowed:
\begin{eqnarray}
W&=&S\left(H^2 \langle X\rangle^2- \langle Y\rangle^2 \right) + N^2H^2 + \langle Y\rangle^2 S^3 \; ,\\
K&=&\left(N\langle Z\rangle + N^*\langle Z\rangle^*\right)^2 + \ldots\,.
\end{eqnarray}
Since we have assumed the fields $X$, $Y$ and $Z$ as heavy and already at their minima, we can neglect them for the dynamics of inflation and just insert their VEVs. Realizing that for a given complex (spatially constant) VEV $\langle Z\rangle$ we can always perform global phase redefinitions of the fields to make $\langle Z\rangle$ real, we see that we recover the simple superpotential and K\"ahler potential of Eqs.~(1) and (2), apart from the additional operator $\langle Y\rangle^2 S^3$. Since $S=0$ during inflation, however, this term has no effect on the inflationary dynamics.

\end{document}